\documentclass[12pt]{article}

\usepackage{amsmath,amssymb}

\newtheorem{theorem}{Theorem}
\newtheorem{proposition}{Proposition}

\newtheorem{definition}{Definition}

\def\derpar#1#2{\frac{\partial{#1}}{\partial{#2}}}
\def\ben{\begin{enumerate}}
\def\een{\end{enumerate}}
\def\bit{\begin{itemize}}
\def\eit{\end{itemize}}
\def\beq{\begin{equation}}
\def\eeq{\end{equation}}
\def\bea{\begin{eqnarray}}
\def\eea{\end{eqnarray}}
\def\beann{\begin{eqnarray*}}
\def\eeann{\end{eqnarray*}}

\def\qed{\ifvmode\removelastskip\fi
{\unskip\nobreak\hfil\penalty50\hbox{}\nobreak\hfil \hbox{\vrule
height1.2ex width1.2ex}\parfillskip=0pt \finalhyphendemerits=0
\par\smallskip}}
\def\vf{\mbox{$\mathfrak X$}}
\def\df{{\mit\Omega}}
\def\Lag{{\cal L}}
\def\FL{{\cal F}\Lag}

\def\Diff{{\rm Diff}}
\def\d{{\rm d}}

\def\Real{\mathbb{R}}

\def\ker{\mathop{\rm ker}\nolimits}
\def\inn{\mathop{i}\nolimits}
\def\Tan{{\rm T}}
\def\Lie{\mathop{\rm L}\nolimits}
\def\Cinfty{\mathrm{C}^\infty}

\title{Some topics concerning the theory of singular dynamical systems}
\author{Narciso Rom\'an-Roy\thanks{{\bf e}-{\it mail}:
  nrr@ma4.upc.edu}\\
\small {\sl  Departamento de Matem\'atica Aplicada IV.} \\
\small {\sl Ed. C-3, Campus Norte UPC. C/ Jordi Girona 1. 08034 Barcelona. Spain}}
\date{ }

\begin{document}

\maketitle

\begin{abstract}
Some subjects related to the geometric theory of
singular dynamical systems are reviewed in this paper.
In particular, the following two matters are considered:
the theory of canonical transformations for presymplectic Hamiltonian
systems, and the Lagrangian and Hamiltonian constraint algorithms
and the time-evolution operator.
\end{abstract}


\bigskip
\vbox{\raggedleft AMS s.\,c.\,(2000): 37J05, 53D05, 53D22, 70H03, 70H05, 70H15, 70H45.
\qquad
PACS (1999): 02.40.Hw, 11.10.Ef, 45.20.Jj}\null

\pagestyle{myheadings}
\markright{\sc N. Rom\'an-Roy,
   \sl Some topics concerning the theory of singular systems}

\section{Introduction}

The aim of this paper is to carry out a brief review 
of several topics concerning
the theory of autonomous singular dynamical systems,
from a geometrical perspective.
In particular, our interest will be focused on two
subjects, namely:
the theory of canonical transformations for singular
systems, and the problem of the
compatibility of the dynamical equations of
Lagrangian and Hamiltonian singular systems;
more precisely, the analysis of the Lagrangian and Hamiltonian constraint
algorithms and their relation.

The article is based on the developments made on the
references \cite{GLR-84} and \cite{CGIR-85}, for the theory of canonical transformations,
and \cite{BGPR-86}, \cite{CLR-87}, \cite{CLR-88},
\cite{MR-92}, \cite{CL-87} and \cite{GP-89}
for aspects related with the constraint algorithms.
Thus, we will refer to these articles for more details
on all these results.

All manifolds are real and $C^\infty$. All
 maps are $C^\infty$. Sum over crossed repeated indices is
 understood. Throughout this paper $\inn(X)\omega$ denotes
the contraction between the vector field $X$ and the differential form $\omega$,
and $\Lie(X)\omega$ the Lie derivative of $\omega$ with respect to the vector field $X$.

\section{Canonical transformations}

\subsection{Presymplectic Locally Hamiltonian Systems}

\begin{definition}
A \emph{presymplectic locally Hamiltonian system} (p.l.h.s.) is a
triad $(M,\omega,\alpha)$, where $(M,\omega)$ is a presymplectic
manifold, and $\alpha\in\ Z^1(M)$ (i.e., it is a closed $1$-form),
which is called a \emph{Hamiltonian form}.
\end{definition}

A p.l.h.s has associated the following equation
$$
\inn(X)\omega=\alpha \quad ,\quad X\in\vf (M)
$$
If $X$ exists, it is called a {\sl presymplectic locally Hamiltonian vector field}
associated to $\alpha$. Nevertheless,
in the best cases, this equation has consistent solutions only in
a submanifold $\jmath_C\colon C\hookrightarrow M$, where
there exist $X\in\vf(M)$, tangent to $C$, such that
\beq
[\inn(X)\omega-\alpha]\vert_C=0
\label{HeqC}
\eeq
Furthermore, the solution $X$ is not unique, in general,
and this non-uniqueness is known as {\sl gauge freedom}.
In general $(C,\omega_{_C}=\jmath_C^*\omega)$ is a presymplectic
manifold which is called the  {\sl final constraint submanifold} (f.c.s.).

The following theorem gives the local structure of p.l.h.s.
(see \cite{CGIR-85}):

\begin{theorem}
 Let $(M,\omega,\alpha)$ be a
p.l.h.s., and $\jmath_C\colon C\hookrightarrow M$ the f.c.s. Then:

1. There are a symplectic manifold $(P,\Omega)$ and a
coisotropic embedding $\imath_C~\colon~C~\hookrightarrow~P$ such that
$\omega_{_C}=\imath_C^*\Omega$. 

2. For every $X\in\vf(M)$,
tangent to $C$, solution to (\ref{HeqC}), there exists a family of
vector fields $\vf(P,C)\subset\vf(P)$ such that, for
every $X_\xi\in\vf(P,C)$, 
(a) $X_\xi$ are tangent to $C$, 
(b) $X_\xi\vert_C=X\vert_C$, and (c)
$X_\xi$ are solutions to the equations
\ $\inn(X_\xi)\Omega=\alpha_P+\xi$,\ 
where $\alpha_P\in Z^1(P)$ satisfies
$\imath_C^*\alpha_P=\jmath_C^*\alpha$, and $\xi\in Z^1(P)$ is any
\emph{first-class constraint form} (i.e., $\imath_C^*\xi=0$, and
the Hamiltonian vector field associated with $\xi$,
$X_\xi\in\vf(P)$, is tangent to $C$).

3. The coisotropic embedding $\imath_C$ and the family $\vf(P,C)$ are
unique, up to an equivalence relation of local symplectomorphisms
reducing to the identity on $C$. 

$(P,C,\Omega)$ is the {\rm coisotropic canonical system} associated to $(M,\omega,\alpha)$.
\end{theorem}

\subsection{Canonical Transformations for p.l.h.s.}

Let  $\vf_{lh}(C)$ be the set of locally Hamiltonian vector fields in
$(C,\omega_C)$; that is,
$\vf_{lh}(C)=\{ X_C\in\vf(C)\ \vert\ \Lie(X_C)\omega_{_C}=0\}$.

\begin{definition}
Let $(M_i,\omega_i,\alpha_i)$ ($i=1,2$) be a p.l.h.s., with f.c.s.
$\jmath_{_{C_i}}\colon C_i\to M_i$, and
$\omega_{_{C_i}}=\jmath_{_{C_i}}^*\omega_i$, such that
$\dim\,M_1=\dim\, M_2$, $\dim\, C_1=\dim\, C_2$, and
${\rm rank}\,\omega_{_{C_1}}={\rm rank}\,\omega_{_{C_2}}$.
A \emph{canonical transformation} between these systems is a pair
$(\Phi,\phi)$, with $\Phi\in\Diff(M_1,M_2)$ and
$\phi\in\Diff(C_1,C_2)$, such that: 

1. $\Phi\circ\jmath_{_{C_1}}=\jmath_{_{C_2}}\circ\phi$;
that is, we have the commutative diagram
$$
\begin{array}{ccc}
M_1 &
\begin{picture}(55,10)(0,0)
\put(22,6){\mbox{$\Phi$}} \put(0,3){\vector(1,0){55}}
\end{picture} &
M_2
\\
\begin{picture}(15,35)(0,0)
\put(12,15){\mbox{$\jmath_{_{C_1}}$}} \put(8,0){\vector(0,1){35}}
\end{picture}
 & &
\begin{picture}(15,35)(0,0)
\put(-8,15){\mbox{$\jmath_{_{C_2}}$}} \put(8,0){\vector(0,1){35}}
\end{picture}
\\
C_1 &
\begin{picture}(55,10)(0,0)
\put(22,8){\mbox{$\phi$}} \put(0,3){\vector(1,0){55}}
\end{picture}
& C_2
\end{array}
$$

2. $\phi_*(\vf_{lh}(C_1))\subset\vf_{lh}(C_2)$.
\end{definition}

The generalization of Lee Hwa Chung's theorem to presymplectic manifolds
allows us to prove that (see \cite{GLR-84}):

\begin{proposition}
Condition 2 is equivalent to saying that there exists $c\in\Real$
such that
\ $\jmath_{C_1}^*(\Phi^*\omega_2-c\omega_1)=
\phi^*\omega_{_{C_2}}-c\omega_{_{C_1}}=0$.
\end{proposition}

$c$ is called the {\sl valence} of the canonical
transformation. So, {\sl univalent canonical transformations} are
the {\sl presymplectomorfisms} between $C_1$ and $C_2$.

Let $(P_i,C_i,\Omega_i)$ be the coisotropic canonical systems
associated with the p.l.h.s $(M_i,\omega_i,\alpha_i)$, $i=1,2$.
The {\sl class of $\phi$} is defined by
$\{\phi\}=\{ \Psi \in \Diff(P_1,P_2) \ \vert\
\Psi~\circ~\imath_{_{C_1}}~=~\imath_{_{C_2}}~\circ~\phi\}$.
So we have the diagram
$$
\begin{array}{ccccccc}
P_1 &
\begin{picture}(10,10)(0,0)
\put(65,6){\mbox{$\Psi$}}
\put(0,3){\vector(1,0){145}}
\end{picture} & & & & &
P_2
\\
& & M_1 &
\begin{picture}(55,10)(0,0)
\put(22,-10){\mbox{$\Phi$}}
\put(0,3){\vector(1,0){55}}
\end{picture} &
M_2 & &
\\
&
\begin{picture}(10,35)(0,0)
\put(-5,18){\mbox{$\imath_{_{C_1}}$}}
\put(20,0){\vector(-1,2){25}}
\end{picture}
&
\begin{picture}(15,35)(0,0)
\put(12,15){\mbox{$\jmath_{_{C_1}}$}}
\put(8,0){\vector(0,1){35}}
\end{picture}
 & &
\begin{picture}(15,35)(0,0)
\put(-8,15){\mbox{$\jmath_{_{C_2}}$}}
\put(8,0){\vector(0,1){35}}
\end{picture}
&
\begin{picture}(10,35)(0,0)
\put(5,18){\mbox{$\imath_{_{C_2}}$}}
\put(-10,0){\vector(1,2){25}}
\end{picture}
&
\\
& & C_1 &
\begin{picture}(55,10)(0,0)
\put(22,8){\mbox{$\phi$}}
\put(0,3){\vector(1,0){55}}
\end{picture}
& C_2 & &
\end{array}
$$
And therefore we have (see \cite{CGIR-85}):

\begin{theorem}
There exists $\psi\in\{\phi\}$ which is a symplectomorphism between 
the symplectic manifolds $(P_1,\Omega_1)$ and $(P_2,\Omega_2)$.
\end{theorem}

As a particular situation,
we can analyze the canonical transformations in a p.l.h.s.
Thus, let $(M,\omega,\alpha)$ be a p.l.h.s., with f.c.s. 
$\jmath_C\colon C\hookrightarrow M$.
Consider the involutive distribution $\ker\,\omega_C$ in $C$, and
assume that the quotient space $\hat C=C/\ker\,\omega_C$ is a manifold
with natural projection $\hat\pi\colon C\to \hat C$
(it is called the {\sl reduced phase space} associated to the p.l.h.s.).
Then the form $\omega_C$ is $\hat\pi$-projectable; hence
there exists $\hat\Omega\in\df^2(\hat C)$ such that \
$\omega_C=\hat\pi^*\hat\Omega$.\
Furthermore, $(\hat C,\hat\Omega)$ is a symplectic manifold,
and we have the following result (see \cite{CGIR-85}):

\begin{proposition}
Every canonical transformation $(\Phi,\phi)$ in  $(M,\omega,\alpha)$
 leaves the distribution $\ker\,\omega_C$ invariant.
As a consequence, there exists a unique $\hat\phi\in\Diff(\hat C)$
such that:

1. $\hat\phi\circ\hat\pi=\hat\pi\circ\phi$.

2. $\hat\phi$ is a symplectomorphism.
\end{proposition}

Some applications of the geometric theory of
p.l.h.s. and their canonical transformations are the following:
the study of canonical transformations for regular autonomous systems
(it includes a new geometrical description of these kinds of systems based on the
{\sl coisotropic embedding theorem}) \cite{CGIR-87},
the analysis of the geometric structure and construction of canonical transformations for
the free relativistic massive particle \cite{CGIR-87},
the discussion about 
time scaling transformations in Hamiltonian dynamics and their
application to celestial mechanics \cite{CIL-88},
and the construction of realizations of symmetry groups for singular systems
and, as an example, the
Poincar\'e realizations for the free relativistic particle \cite{RR-91}.

\section{Constraints and the Evolution Operator}

\subsection{Lagrangian Dynamical Systems}

(See \cite{Ca-90} for details).

Let $Q$ be a $n$-dimensional differential manifold
which constitutes the {\sl configuration space} of a dynamical system.
Its tangent and cotangent bundles,
$\tau_Q\colon \Tan Q\to Q$\ and $\pi_Q\colon \Tan^*Q\to Q$,
are the {(\sl phase spaces of velocities and momenta} of the system.

A {\sl Lagrangian dynamical system} is a couple $(\Tan Q,\Lag)$, where
$\Lag\in\Cinfty(\Tan Q)$ is the {\sl Lagrangian function} of the system.
Using the canonical elements of $\Tan Q$,
the vertical endomorphism ${\cal S}\in{\mathfrak T}^1_1(\Tan Q)$ and the
Liouville vector field $\Delta\in\vf(\Tan Q)$,
we can define the {\sl Lagrangian $2$-form}
\ $\omega_\Lag=-\d(\d\Lag\circ{\cal S})$,\
and the {\sl Lagrangian energy function}
\ $E_\Lag=\Delta(\Lag)-\Lag$.
Moreover, we define the {\sl Legendre transformation} associated with $\Lag$,
\ $\FL\colon\Tan Q\to\Tan^*Q$,\
as the fiber derivative of the Lagrangian.

$\Lag$ is a {\sl singular Lagrangian}
if $\omega_\Lag$ is a presymplectic form 
(which is assumed to have constant rank) or, what is equivalent,
$\FL$ is no longer a local diffeomorphism.
In particular, $\Lag$ is an {\sl almost-regular Lagrangian}
 if: (i) $M_0=\FL(\Tan Q)$
 is a closed submanifold of $\Tan^*Q$,
 (ii) ${\cal F}\Lag$ is a submersion onto $M_0$, and
 (iii) for every ${\rm p}\in\Tan Q$, the fibres
 ${\cal F}\Lag^{-1}({\cal F}\Lag (\rm p))$
 are connected submanifolds of $\Tan Q$.
Then $(\Tan Q,\Lag)$ is an {\sl almost-regular Lagrangian system}.

For almost-regular Lagrangian systems, 
$(\Tan Q,\Omega_\Lag,\d E_\Lag)$ is a p.l.h.s.,
and we have the so-called {\sl Lagrangian dynamical equation}
\beq
\inn(\Gamma_\Lag)\omega_\Lag=\d E_\Lag
\label{ELeq}
\eeq
Variational considerations lead us to impose that,
solutions $\Gamma_\Lag\in\vf(\Tan Q)$ to (\ref{ELeq}) must be
{\sl second-order differential equations} ({\sc sode});
that is, holonomic vector fields in $\Tan Q$.
Geometrically this means that
\beq
{\cal S}(\Gamma_\Lag)=\Delta
\label{sode}
 \eeq
For singular Lagrangians this does not hold
in general, and (\ref{sode}) must be imposed as an additional condition
({\sl {\sc sode}-condition}).
Integral curves of vector fields satisfying (\ref{ELeq})
and (\ref{sode}) are solutions to the Euler-Lagrange equations.

The {\sl Lagrangian problem} consists in finding a submanifold
$\jmath_{_{S_f}}\colon S_f\hookrightarrow\Tan Q$,
and $\Gamma_\Lag\in\vf (\Tan Q)$, tangent to $S_f$, such that
$$
[\inn(\Gamma_\Lag)\omega_\Lag-dE_\Lag]\vert_{_{S_f}}=0\quad ,\quad
[{\cal S}(\Gamma_\Lag)-\Delta]\vert_{_{S_f}}=0
$$

Now, if $\FL_0\colon\Tan Q\to M_0$ is the restriction of $\FL$ to $M_0$,
we have that $\omega_\Lag$ and $E_\Lag$ are $\FL_0$-projectable: there
exist $\omega_0\in\df^2(M_0)$, and $h_0\in\Cinfty(M_0)$ such that
$\omega_\Lag=\FL_0^*\omega_0$, $E_\Lag=\FL_0^*h_0$.
Then, $(M_0,\omega_0,\d h_0)$ is a p.l.h.s. which is called
the {\sl canonical Hamiltonian system} associated with
the Lagrangian system $(\Tan Q,\omega_\Lag,\d E_\Lag)$
So we have the {\sl Hamiltonian dynamical equation}
$$
\inn(X_0)\omega_0=dh_0 \quad ;\quad
X_0\in\vf(M_0)
$$
and the {\sl Hamiltonian problem} consists in finding
a submanifold $\jmath_{_{M_f}}\colon M_f\hookrightarrow M_0$ and
$X_0\in\vf(M_0)$ tangent to $M_f$ such that
$$
[\inn(X_0)\omega_0-dh_0]\vert_{_{M_f}}=0
$$

\subsection{Constraint Algorithms}

In order to solve the Hamiltonian problem stated for an
almost-regular system different kinds of Hamiltonian constraint algorithms
were developed.
The first was the local-coordinate {\sl Dirac constraint algorithm} \cite{Di-64},
but there were also geometric algorithms:
the {\sl Presymplectic Constraint Algorithm} (PCA) of {\it Gotay,
Nester, Hinds} \cite{GNH-78}, and others by
{\it Marmo, Tulczyjew} et al. \cite{MMT-97}, \cite{MT-78}, etc.
All of them give a sequence of submanifolds which,
in the best cases, stabilizes giving the f.c.s.:
$\Tan^*Q\hookleftarrow M_0\hookleftarrow M_1\hookleftarrow \ldots\hookleftarrow M_f$.

For the Lagrangian problem, the first attempt was not to consider
the {\sc sode}-problem  (\ref{sode}), and then develop a Lagrangian constraint algorithm
by simply applying the P.C.A. to the Lagrangian dynamical equation (\ref{ELeq}),
obtaining a sequence of submanifolds
$\Tan Q\hookleftarrow P_1\hookleftarrow \ldots\hookleftarrow P_f$\
\cite{GN-80a}. The {\sc sode} problem is studied later in \cite{GN-80},
obtaining a submanifold $S_f$ which solves the lagrangian problem, but
which is not defined by constraints and is not maximal.
Later, {\it Kamimura} \cite{Ka-82} and {\it Batlle, Gomis, Pons, Rom\'an-Roy}
\cite{BGPR-86} developed local-coordinate Lagrangian constraint algorithms
in which the Lagrangian dynamical equation (\ref{ELeq}) and the 
{\sc sode}-condition (\ref{sode}) were both considered at the same time.
The f.c.s. $S_f$ obtained at the end of the corresponding sequence,
$\Tan^*Q\hookleftarrow S_1\hookleftarrow \ldots\hookleftarrow S_f$,
is maximal and is defined by constraints.
The relation between the aforementioned sequences of submanifolds
is explained in the following diagram
\beq
\begin{array}{cccccccc}
\Tan Q  & \hookleftarrow & P_1 & \hookleftarrow & \ldots & \hookleftarrow & P_f & \\
& & & \begin{picture}(5,10)(0,0)
\put(0,0){\vector(-1,1){10}}
\end{picture} & & & &
\begin{picture}(5,10)(0,0)
\put(0,0){\vector(-1,1){10}}
\end{picture}\\
& & & S_1 & \hookleftarrow & \ldots & \hookleftarrow & S_f
\label{sequence}
\end{array}
\eeq
In particular, the submanifolds
$P_i$, $i=1\ldots f$, are defined by constraints that
can be expressed as {\sl $\FL$-projectable} functions which give
all the Hamiltonian constraints, and are
related with the {\sl first-class Hamiltonian constraints}.
Futhermore, the submanifolds $S_i$, $i=1\ldots f$, are defined by adding constraints that
are not {\sl $\FL$-projectable},
and they are related with the {\sl second-class Hamiltonian constraints}.

The remaining question was
how to describe geometrically the submanifolds $S_i$ and their properties.
First, this problem was solved in \cite{CLR-87}, \cite{CLR-88} for $S_1$,
i.e.; the submanifold of compatibility conditions for (\ref{ELeq}) and (\ref{sode}),
obtaining as the main result that:

\begin{theorem}
For every {\sc sode} $\Gamma\in\vf(\Tan Q)$, we have:
$$
S_1=\{{\rm p}\in\Tan Q\ \vert\
[\inn(Z)(\inn(\Gamma)\omega_\Lag-\d E_\Lag)]({\rm p})=0,\ \forall Z\in{\cal M}\}
$$
where ${\cal M}=\{ Z\in\vf(\Tan Q)\ \vert\
{\cal S}(Z)\in\ker\,\FL_*=\ker\,\omega_\Lag\cap\vf^{{\rm V}(\tau_Q)}(\Tan Q)\}$.

In particular, for every $Z\in\ker\,\omega_\Lag\subset{\cal M}$,\
$\inn(Z)(\inn(\Gamma)\omega_\Lag-\d E_\Lag)=\inn(Z)(\d E_\Lag)$
define the submanifold $P_1$ where (\ref{ELeq}) is compatible.
They are called {\rm dynamical constraints},
and are related to the existence of primary first-class Hamiltonian constraints.

For every $Z\not\in\ker\,\omega_\Lag$, $Z\in{\cal M}$, these
functions are called {\rm {\sc sode}-constraints}, and are
related to the existence of primary second-class Hamiltonian constraints.
\end{theorem}

For $S_i$, $i=2\ldots f$, the problem was studied in \cite{MR-92},
by imposing tangency conditions for solutions to (\ref{ELeq}) and (\ref{sode}).
The main results are the following:
there are two kinds of Lagrangian constraints defining every $S_i$, $i=1,\ldots f$:

- {\sl Dynamical constraints}, which define the submanifolds $P_i$, $i=1,\ldots f$,
in the sequence (\ref{sequence}). They are related with the solutions to the eq. (\ref{ELeq}),
and all of them can be expressed as $\FL$-projectable functions
which give all the Hamiltonian constraints.

- {\sl {\sc sode} (non-dynamical) constraints},
coming from the {\sc sode}-condition (\ref{sode}).
They are not \emph{$\FL$-projectable}.

Hence, the relation among the sequences of submanifolds
in the Lagrangian and Hamiltonian formalisms is given in
the following diagram
\beann
\begin{array}{cccccccccc}
& & \Tan Q  & \hookleftarrow & P_1 & \hookleftarrow & \ldots & \hookleftarrow & P_f & \\
& & & & & \begin{picture}(5,10)(0,0)
\put(0,0){\vector(-1,1){10}}
\end{picture} & & & &
\begin{picture}(5,10)(0,0)
\put(0,0){\vector(-1,1){10}}
\end{picture}\\
& & & & & S_1 & \hookleftarrow & \ldots & \hookleftarrow & S_f\\
& \begin{picture}(2,25)(0,0)
\put(20,50){\vector(-1,-1){48}}
\put(-27,25){\mbox{$\FL$}}
\end{picture} &
\begin{picture}(2,25)(0,0)
\put(0,50){\vector(0,-1){50}}
\put(5,15){\mbox{$\FL_0$}}
\end{picture} & &
\begin{picture}(2,25)(0,0)
\put(0,50){\vector(0,-1){50}}
\end{picture}
& \begin{picture}(2,25)(0,0)
\put(-2,25){\vector(-1,-2){12}}
\end{picture}
& & &
\begin{picture}(2,25)(0,0)
\put(0,50){\vector(0,-1){50}}
\end{picture}
&
\begin{picture}(2,25)(0,0)
\put(-2,25){\vector(-1,-2){12}}
\end{picture} \\
\Tan^*Q & \hookleftarrow & M_0 & \hookleftarrow &
M_1 & \hookleftarrow & \ldots & \hookleftarrow & M_f &
\end{array}
\eeann
Furthermore, the tangency conditions for dynamical constraints give
dynamical and {\sc sode} constraints, while the tangency conditions for 
{\sc sode}-constraints remove gauge degrees of freedom
and do not give new constraints.

\subsection{The Time Evolution Operator $K$}

As a final remark, the complete relation between Hamiltonian and Lagrangian constraints
is given by the so-called {\sl time evolution operator} $K$,
which relates Lagrangian and Hamiltonian constraints, and also
solutions to the Lagrangian and Hamiltonian problems.
(It is also known by the name of the {\sl relative Hamiltonian vector field}
\cite{PV-00}). It was introduced and studied for the first time by {\it J. Gomis} et al.
\cite{BGPR-86}, developing some previous ideas of {\it K. Kamimura} \cite{Ka-82}.
They give the local-coordinate definition of this operator,
whose expression in coordinates is
$$
K=v^A\left(\derpar{}{q^A}\circ{\cal F}\Lag\right)+
\derpar{\Lag}{q^A}\left(\derpar{}{p^A}\circ{\cal F}\Lag\right)
$$
The intrinsic definition and the geometric study of its properties
was carried out independently in \cite{CL-87} and \cite{GP-89}.
In the first work, $K$ was defined using the 
{\it Skinner-Rusk} {\sl unified formalism} in $\Tan Q\oplus\Tan^*Q$.
In the second article, the concept of {\sl section along a map}
plays the crucial role, and so we have:

\begin{definition}
Let $(\Tan Q,\omega_\Lag,E_\Lag)$ be a Lagrangian system,
and $\Omega\in\df^2(\Tan^*Q)$ the canonical form.
The {\sl time-evolution operator $K$} associated
with $(\Tan Q,\omega_\Lag,{\rm E}_\Lag)$ is a map
$K\colon \Tan Q\longrightarrow\Tan\Tan^*Q$
verifying the following conditions:

1. ({\sl Structural condition}):
 $K$ is vector field along ${\cal F}\Lag$,\quad
$\pi_{\Tan^*Q}\circ K={\cal F}\Lag$.

2. ({\sl Dynamical condition}):\quad
${\cal F}\Lag^*[\inn(K)(\Omega\circ{\cal F}\Lag)]=\d E_\Lag$.

3. ({\sl {\sc sode} condition}):\quad
 $\Tan\tau_Q\circ K={\rm Id}_{\Tan Q}$.
$$
\begin{array}{ccc}
 \Tan Q &
\begin{picture}(30,10)(0,0)
\put(10,8){\mbox{$\Tan\tau_Q$}}
\put(38,3){\vector(-1,0){38}}
\end{picture}
& \Tan\Tan^*Q
\\
\begin{picture}(10,30)(0,0)
\put(-24,12){\mbox{${\rm Id}_{\Tan Q}$}}
\put(3,30){\vector(0,-1){30}}
\end{picture}
 &
\begin{picture}(30,30)(0,0)
\put(8,19){\mbox{$K$}}
\put(0,0){\vector(1,1){30}}
\end{picture}
&
\begin{picture}(10,30)(0,0)
\put(8,12){\mbox{$\pi_{\Tan^*Q}$}}
\put(3,30){\vector(0,-1){30}}
\end{picture}
\\
\Tan Q &
\begin{picture}(30,10)(0,0)
\put(10,6){\mbox{${\FL}$}}
\put(0,3){\vector(1,0){38}}
\end{picture}
& \Tan^*Q
\end{array}
$$
\end{definition}

Then, the relation between Lagrangian and Hamiltonian constraints
is established as follows (see \cite{BGPR-86}, \cite{CL-87}, \cite{GP-89}):

\begin{proposition}
If $\xi\in\Cinfty(\Tan^*Q)$ is a $i$th-generation Hamiltonian constraint,
then $\Lie(K)\xi$ is a $(i+1)$th-generation Lagrangian constraint.

In particular, if $\xi$ is a {\rm first-class constraint}
(resp. a {\rm second-class constraint}) for $M_f$,
then $\Lie(K)\xi$ is a dynamical constraint
(resp. a {\sc sode} constraint).
\end{proposition}

The time-evolution operator has also been used for studying
different kinds of problems concerning singular systems.
For instance, the operator $K$ has been extended for
analyzing higher-order singular dynamical systems
\cite{BGPR-88}, \cite{CL-92}, \cite{GPR-91}, \cite{GPR-92}.
It is also used for treating constrained systems in
general ({\sl linearly singular systems}) \cite{GP-92}.
It has been defined and its properties studied for
non-autonomous dynamical systems \cite{CFM-95}.
Finally, $K$ has been applied for analyzing
gauge symmetries and other structures for singular systems
\cite{GP-88}, \cite{GP-2001}.
Furthermore, sections along maps in general are analyzed and used 
in different kinds of physical and geometrical problems in
\cite{Ca-96}, \cite{CMS-92}, \cite{CMS-93}.

\section{Discussion and outlook}

Some of the previous problems have been studied in the sphere of
first-order classical field theories, specially their
{\sl multisymplectic formalism} \cite{CCI-91}. So,
a geometric constraint algorithm has recently been completed
for Lagrangian and Hamiltonian singular field theories \cite{LMMMR-05},
and the definition and properties of the operator $K$ have
been carried out for field theories \cite{EMMR-03}, \cite{RRS-05}.

Other potentially interesting topics could be the
generalization of some of the above results;
such as: to study the local structure of pre-multisymplectic
Hamiltonian field theories
(previous generalization of the {\sl coisotropic embedding
theorem} for premultisymplectic manifolds);
the study of canonical transformations for Hamiltonian field theories
({\sl multisymplectomorphisms} and {\sl pre-multisymplectomorphisms\/}), and
the application of the operator $K$ to analyze the relation between
Lagrangian and Hamiltonian constraints of singular field theories
(which could require prior development of the non-covariant formulation,
i.e., {\sl space-time splitting}, of field theories).

\subsection*{Acknowledgments}

I acknowledge the financial support of
{\sl Ministerio de Educaci\'on y Ciencia}, project
MTM2004-7832.
I am indebted to the organizers of the 
{\sl International Workshop on Groups, Geometry and Physics}
for inviting me to participate on it.
I wish also to thank Mr. Jeff Palmer for his
assistance in preparing the English version of the manuscript.

\end{document}